# NMR study of Ni$_{50+x}$Ti$_{50-x}$ Strain-Glass

Rui Li,[1] Jacob Santiago,[1] Daniel Salas,[2] Ibrahim Karaman,[2] and Joseph H. Ross Jr[1, 2]

[1]*Department of Physics and Astronomy, Texas A&M University, College Station, Texas 77843*
[2]*Department of Materials Science and Engineering,*
*Texas A&M University, College Station, Texas 77843*
(Dated: July 27, 2023)

We studied Ni$_{50+x}$Ti$_{50-x}$ with compositions up to $x = 2$, performing $^{47}$Ti and $^{49}$Ti nuclear magnetic resonance (NMR) measurements from 4 K to 400 K. For large $x$ in this system, a strain-glass appears in which frozen ferroelastic nano-domains replace the displacive martensite structural transition. Here we demonstrate that NMR can provide an extremely effective probe of the strain glass freezing process, with large changes in NMR line-shape due to the effects of random strains which become motionally narrowed at high temperatures. At the same time with high-resolution x-ray diffraction we confirm the lack of structural changes in $x \geq 1.2$ samples, while we show that there is little change in the electronic behavior across the strain glass freezing temperature. NMR spin-lattice relaxation time ($T_1$) measurements provide a further measure of the dynamics of the freezing process, and indicate a predominantly thermally activated behavior both above and below the strain-glass freezing temperature. We show that the strain glass results are consistent with a very small density of critically divergent domains undergoing a Vogel-Fulcher-type freezing process, coexisting with domains exhibiting faster dynamics and stronger pinning.

## I. INTRODUCTION

Shape memory alloys (SMAs) have been attractive for decades due to their potential applications in many technological and scientific areas [1]. Among these, NiTi has been one of the most promising materials due to its applications in robotics, biomedical and aerospace, due to its ability to recover its original shape upon deformation as a result of the reversible martensitic transformation. Stoichiometric NiTi undergoes a first-order phase transition from the high-temperature cubic (CsCl) austenite B2 phase (space group: $Pm\bar{3}m$) into the monoclinic martensite B19' phase (space group: $P2_1/m$) upon temperature decreases (see Fig. 1).

Extensive experimental and theoretical efforts have been made in recent years to tailor martensitic transitions for controlled strain recovery [2–5]. One approach is to turn the abrupt, first-order transformation into a broadly diffused, nearly continuous, strain glass transition by adding defects. With sufficient surplus Ni content, the martensite transition can be completely suppressed and the Ni$_{50+x}$Ti$_{50-x}$ system becomes a strain glass, characterized by the formation of nanodomains of large local strains (i.e., a strain glass state) through a freezing transition and exhibits a continuous transformation behavior upon cooling/heating cycling. The strain glass appears to have no martensitic transformation in calorimetry or x-ray diffraction. However, below a well-defined freezing temperature, frequency dispersion of the storage modulus appears in dynamic mechanical analysis [6–9].

The strain glass is believed to be a kind of ferroic glass, which also includes spin glass and relaxor ferroelectrics [10, 11]. Although spin glasses and relaxor ferroelectrics were discovered many decades ago, strain glasses were first identified in 2005 [6] following earlier theoretical work [12, 13] showing that glassy behavior may arise from premartensitic phases above the transformation temperature. These ferroic glasses are defined by disorder of the corresponding order parameters that characterize the associated transformations, i.e., strain in strain glasses, spin in spin glasses, and polarization in relaxor ferroelectrics, thus unlike other structural glasses these glasses emerge from an ordered crystalline state.

Nuclear magnetic resonance (NMR) spectroscopy provides a probe of this behavior on an atomic scale, revealing both static and dynamic information. Here, for the first time, we have applied this method as a probe of strain glass dynamics. We performed NMR studies of Ni$_{50+x}$Ti$_{50-x}$ samples from $x = 0.1$ to 2.0, spanning the range from the martensitically transforming to non-transforming strain glass compositions. In addition, we performed high-resolution x-ray diffraction (XRD) to characterize changes in crystal structure and density functional theory (DFT) calculations to model the electronic structure. As a result we provide information about the thermally activated changes in dynamics that characterize the freezing process.

## II. EXPERIMENTAL METHODS

### A. Sample preparation

Ni$_{50+x}$Ti$_{50-x}$ (in at.%) samples include $x = 0.1$ powder fabricated via electrode induction-melting gas atomization (EIGA, from Carpenter Technology Corp.) and $x = 1.2$ bulk material was acquired from Fort Wayne Metals and then gas-atomized by Nanoval GmbH & Co. KG. An FEI Quanta 600 FE scanning electron microscope was used to investigate the powder morphology and size distribution. Spherical particles with smooth surfaces and no hollow particles or satellite particles were observed. Mean diameters are close to 29 $\mu$m for $x = 0.1$ and 20 $\mu$m for $x = 1.2$. Both types of powders were wrapped with thin Ta foil and sealed on quartz tubes under a low-pressure high-purity Ar atmosphere. They were homogenized at 1223 K during 24 h and quenched in room temperature water. Powders suffered mild sintering and area-limited oxidation. Affected areas were disposed of and the rest was easily re-powdered mechanically. For NMR, the powders were sealed in vials with a specimen mass of around 2500 mg. A bulk billet with the nominal $x = 2.0$ content was prepared using vacuum induction melting and then hot-rolled at 1173 K in multiple steps to a 6.35 mm thick plate. Foils with the dimension of 20 mm × 10 mm × 300 $\mu$m were sliced from this plate using wire electrical discharge machining (wire EDM). Following, the foils were sealed on quartz tubes under a low-pressure high-purity Ar atmosphere, heat treated at 1173 K for 1 h for homogenization, and then quenched in room temperature water. The NMR testing samples were prepared by stacking 10 of these foils and separating them using thin insulating spacers to prevent induced currents.

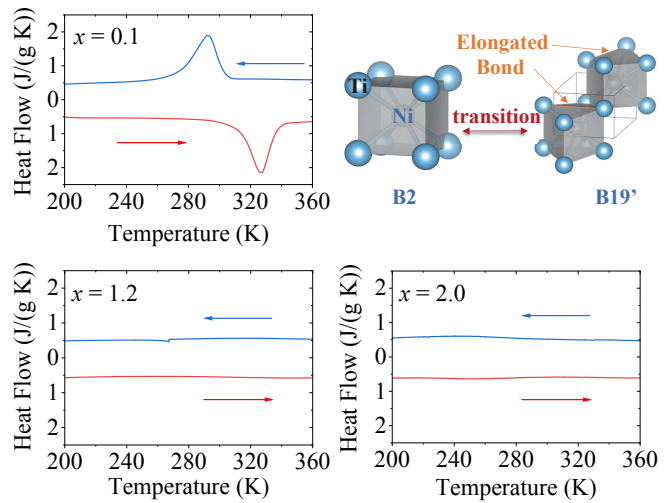

FIG. 1. DSC calorimetry results for three $Ni_{50+x}Ti_{50-x}$ samples as shown, with arrows indicating the cooling/heating direction. Latent heat peaks show the well-defined martensitic transition for $x = 0.1$, while no transition is seen for $x = 1.2$ and 2.0 samples. Upper right: B2 and B19' structures associated with the martensitic transition.

## B. Differential Scanning Calorimetry (DSC)

Differential scanning calorimetry (DSC) was performed to characterize the transition types using a TA Instruments DSC Q2000 with a cooling/heating rate of 10 K/min. For $x = 0.1$, the first-order martensitic transformation gives the obvious exothermic and endothermic peaks (Fig. 1) due to the latent heat of transformation. These peaks show thermal hysteresis, as is typical for this material. By contrast, in $x = 1.2$ and 2.0 samples, the strain glass transition shows no obvious exothermic and endothermic peaks upon cooling and heating, implying no transformation.

## C. X-ray Diffraction (XRD)

High-resolution powder X-ray diffraction (XRD) spectra were obtained at the Advanced Photon Source at Argonne National Laboratory, with excitation wavelength 0.459012 Å. These included $x = 0.1$ and 1.2 samples from the powders processed for NMR, while for $x = 2.0$, one of the foils was wire EDM cut into thin strips which were etched and placed in a Mylar capillary for XRD, with the excitation position translated along the foils during measurement to ensure adequate powder statistics. For $x = 0.1$, measurements at $T = 100$ and 400 K bracketed the martensitic transformation, while measurements were taken at 100, 200, and 295 K for $x = 1.2$ and 2.0 to sample changes associated with strain-glass freezing. GSAS-II software [14] was used for Rietveld refinement of the resulting patterns.

## D. Nuclear Magnetic Resonance (NMR)

$^{47}$Ti ($I = 5/2$) and $^{49}$Ti ($I = 7/2$) NMR measurements were performed using a custom-built NMR spectrometer at a Larmor frequency $\nu_L(Ti) = 21.5$ MHz over temperatures from 4 K to 400 K. These two NMR isotopes have nearly identical gyromagnetic ratios ($\gamma$) so that their spectra overlap. Thus the results shown here are combined NMR spectra, obtained using a standard spin-echo sequence by superposing fast Fourier transformation spectra at a sequence of frequencies. For all displayed spectra, the zero shift position is that of the $^{47}$Ti NMR standard (TiCl$_4$ liquid) calibrated using SrTiO$_3$ as a secondary NMR reference with its -843 ppm NMR shift [15] relative to TiCl$_4$.

As is typical for $I > 1/2$ half-integer nuclei, the largest NMR shift is an electric quadrupole shift [16], parameterized by $\nu_Q = \frac{3eQV_{zz}}{2I(2I-1)h}$ and $\eta = (V_{xx} - V_{yy})/V_{zz}$ where $V_{jj}$ are electric field gradients (EFGs) and $Q$ is the nuclear quadrupole moment. In addition the magnetic shift ($K$) combines the Knight shift due to susceptibility of the metallic electrons with the chemical shift (conventionally $\delta$) due to susceptibility of filled orbitals. These combine to give a relative shift $K = (\nu - \nu_L)/\nu_L$. The central transition ($1/2 \leftrightarrow -1/2$), displayed for most plotted spectra, is strongest and was used to measure the spin-lattice relaxation time ($T_1$). In addition, the satellite ($\pm 7/2 \leftrightarrow \pm 5/2$, $\pm 5/2 \leftrightarrow \pm 3/2$ and $\pm 3/2 \leftrightarrow \pm 1/2$) spectra were measured by echo integration.

### E. Density Functional Theory (DFT)

DFT calculations were performed using the linearized augmented plane-wave + local orbitals (LAPW+lo) method as implemented in the WIEN2k code [17, 18]. Calculations utilized the Perdew-Burke-Ernzerhof (PBE) functional [19] with a mesh of 1000 irreducible k-points, and a cutoff parameter $k_{max} = 7/RMT$ inside the interstitial region for plane wave expansions. These included electric field gradient (EFG) calculations, a sensitive probe of local strain distortions. Similar calculation methods are typically expected to give close agreement with observed EFGs, for example within about 10% for comparable intermetallics [20].

## III. XRD RESULTS

XRD results are shown in Fig. 2. Small reflections due to Ti$_2$Ni ($< 1\%$ per mole formula unit) were seen in all samples, which are oxygen-stabilized precipitates typically found in this material [21]. Otherwise refinements show only B2 and B19′ structures [22], with the cubic B2 phase found for the strain glass at all temperatures, similar to what has been indicated for other strain glasses [23–25].

For $x = 0.1$, the 100 K results fit the expected B19′ monoclinic structure [Fig. 2 (b)]. The broad feature in the difference curve near 12° may possibly correspond to a small nanostructured R-phase superstructure component (see supplemental material [26].), however, there is no evidence for a significant amount of other phases present [38]. Unlike the $x = 0.1$ B2 phase, the peak profiles for the other two samples could not be fitted using the pseudo-Voigt profile of GSAS-II, having instead broader tails flanking each sharp reflection. These were successfully refined in a two-phase model, with sharp reflections combined with those of a second identical B2 phase with large microstrain: for example the highly strained reflections for $x = 1.2$ refined to 26% phase fraction, with isotropic microstrain broadening modeled within the GSAS-II package by a $\tan(\theta)$ dependent adjustment of the pseudo-Voigt parameters. The highly strained regions could correspond to the coherent nanodomains [5, 39], although this model is intended here mainly as a way to quantify changes in the spectra versus temperature. The fits have almost no temperature dependence. More details, and spectra for other temperatures, are shown in the supplemental material [26].

## IV. NMR RESULTS

### A. Lineshapes

Fig. 3 displays central transition NMR spectra for the three samples versus temperature. The $x = 0.1$ lineshape exhibits a large change near room temperature due to the martensitic transition, above which one can clearly see the separate $^{47}$Ti and $^{49}$Ti NMR peaks. This is due to the absence of EFG's, and hence narrowing of the quadrupole spectra, in the cubic B2 phase, consistent with previously reported results [40]. For $x = 1.2$ and 2.0, line broadening sets in (Fig. 3(d)) at temperatures consistent with the onset of strain glass freezing, despite the absence of changes in XRD. Above this temperature, where electron microscopy results imply an unfrozen strain glass [5], the NMR results are consistent with a broadened B2 phase.

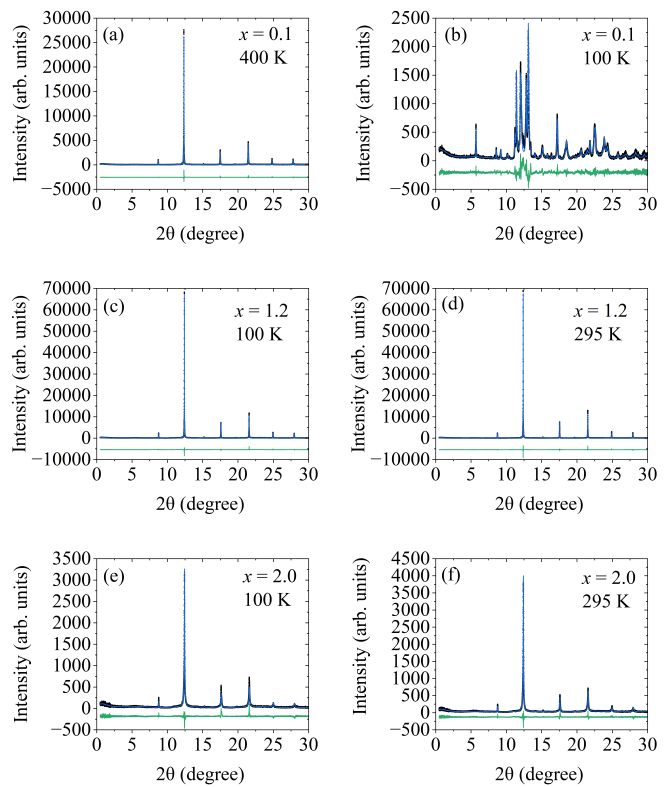

FIG. 2. X-ray powder diffraction data for samples and temperatures indicated. Refined spectra (blue curves) and difference curves (green) also shown. Data residual: Rw = 8.9%, 15.9%, 9.5%, 9.7%, 15.2%, 14.1% for (a-f) respectively. Small Ti$_2$Ni reflections refined to 0.8% / 0.6% / 0.5% per mole formula-unit for $x = 0.1$ / 1.2 / 2.0, with otherwise only B19′ phase in (b), and B2 phase in all other refined spectra.

### B. NMR shift

NMR shifts versus temperature for the martensitically transforming $x = 0.1$ are shown in Fig. 4. In the martensite phase, the peak intensity positions of the broadened spectra are plotted. The wide-line spectrum measured at 4 K (upper inset) has large tails dominated by electric quadrupole interactions. Least-squares fitting using a custom lineshape program yielded the parameters $\eta = 0.60$, and $^{47}\nu_Q = 775$ kHz / $^{49}\nu_Q = 300$ kHz (for $^{47}$Ti / $^{49}$Ti, the ratio determined by the corresponding nuclear parameters [41]). These are in close agreement with the previously reported [40]: $^{47}\nu_Q = 750$ kHz, $\eta = 0.60$. As a result, we obtained $K = 2820$ ppm at 4 K, slightly larger than the 2700 ppm (renormalized for TiCl$_4$ standard) previously obtained [40] for the NiTi martensite. For the austenite phase, we obtain $K = 2973$ ppm at 330 K.

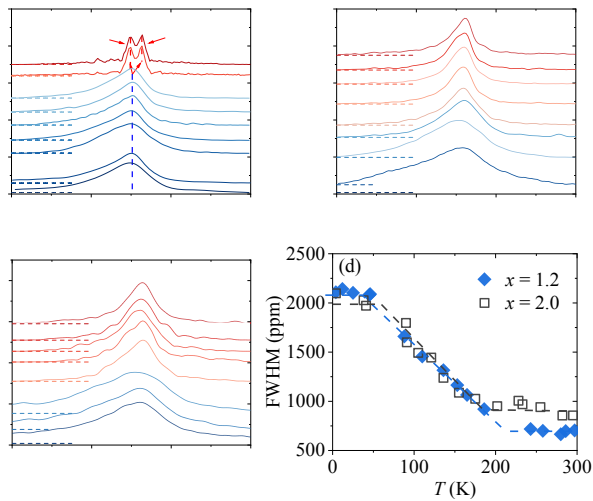

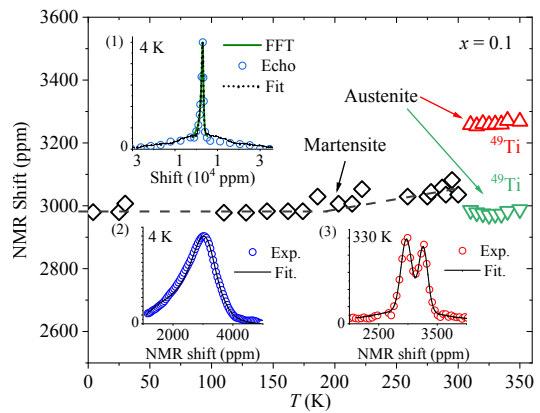

FIG. 4. Warming curve NMR shifts for $x = 0.1$ sample, relative to $^{47}$Ti standard. Martensite data points are maximum intensity positions of the superposed spectra of the two nuclei. (1): 4 K broad-line spin echo spectrum with FFT central transition, and fitted curve described in the text. (2,3): Fitted 4 K and 330 K central-transition FFT spectra.

FIG. 3. (a-c) $^{47}$Ti, $^{49}$Ti NMR spectra for Ni$_{50+x}$Ti$_{50-x}$ samples over temperatures from 4 K to 400 K, normalized to the same height. For each spectrum, the baseline (dashed line) is positioned at its measurement temperature. (d) Linewidths for strain glass compositions as measured by the full width at half maximum (FWHM)

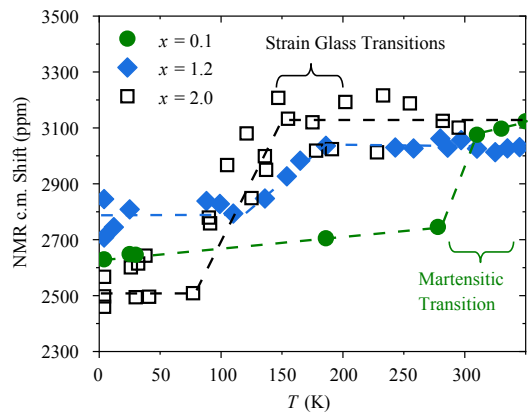

FIG. 5. NMR center of mass (c.m.) shift for $x = 1.2$ and 2.0 strain glass samples, as well as c.m. for $x = 0.1$ martensite transition sample. Dashed lines are guide to the eye.

For the strain glass samples, central-transition center of mass (c.m.) shifts are shown in Fig. 5. The c.m. is defined as the intensity-weighted average shift. For these results, the spectra were fitted using two Gaussian peaks, from which the c.m. shift was calculated. Changes in c.m. correspond to the appearance of enhanced line broadening [Fig. 3(d)] and are due largely to the development of nonzero EFGs at the Ti sites, which induce a negative c.m. shift for the central transition due to second-order electric quadrupole effects [42]. This indicates substantial distortion away from the local cubic environment of the B2 phase. For comparison, the c.m. for $x = 0.1$ sample is also shown in Fig. 5, clearly indicating the martensite transition near room temperature. The frozen strain-glass spectra can be seen to develop asymmetric lineshapes [Fig. 6(a-b)] accompanying the changes in c.m. shift.

As a quantitative measure of the local-scale asymmetry of the frozen strain-glass, we analyzed the low-temperature wide-line spectra [Fig. 6(c-d)] in a similar manner as described above for $x = 0.1$. Making the simplifying assumption that the strain glass broadened spectra could be described by a single set of $^{47}$Ti NMR parameters, the numerical fits shown yielded $^{47}\nu_Q = 1100$ (1000) kHz and $\eta = 0.70$ (0.75), for $x = 1.2$ (2.0). These larger magnitude EFG's are indicative of atomic-scale asymmetry comparable to that of the coherently distorted B19′ phase. The magnetic shifts $K$ are 2950 and 3140 ppm respectively.

## C. Spin-lattice relaxation

The NMR spin-lattice relaxation time ($T_1$) was measured using a spin echo inversion-recovery sequence and fitted with a single exponential curve: $M(t) = \left(1 - Ae^{-\frac{t}{T_1}}\right)M(\infty)$. Here, $M(t)$ is the measured signal and $t$ is the recovery time. This yielded the $1/T_1$ results shown in Fig. 7. Due to the small signal in the stacked-foil sample, we did not perform similar measurements versus temperature for $x = 2.0$.

For both samples $(T_1T)^{-1} = $ constant at low temperatures, a characteristic result for metals referred to as the Korringa process [16, 42] due to magnetic interactions with conduction electrons. The increase in relaxation rate at higher temperatures shows the influence of atomic fluctuations, which contribute to $1/T_1$ via electric quadrupole hyperfine interactions. From the temperature dependence and large magnitude of the relaxation rate, particularly in the strain glass, we can rule out phonon-related dynamics [43, 44]. For $x = 0.1$, there is a significant further increase in $1/T_1$ at the onset of the phase transition; A narrow $1/T_1$ peak is commonly observed near phase transformations [45] due to critical fluctuations. In relaxor ferroelectrics, in many ways the analogs of strain glasses, a similar peak in $1/T_1$ is often observed near the freezing temperature [46], as will be discussed in more detail below. Here, there is no sign of a $1/T_1$ peak at the strain glass freezing temperature; Rather, for both $x = 0.1$ and $x = 1.2$ the results can be interpreted in terms of thermally activated fluctuations, gradually increasing versus $T$ and greatly enhanced in the strain glass sample.

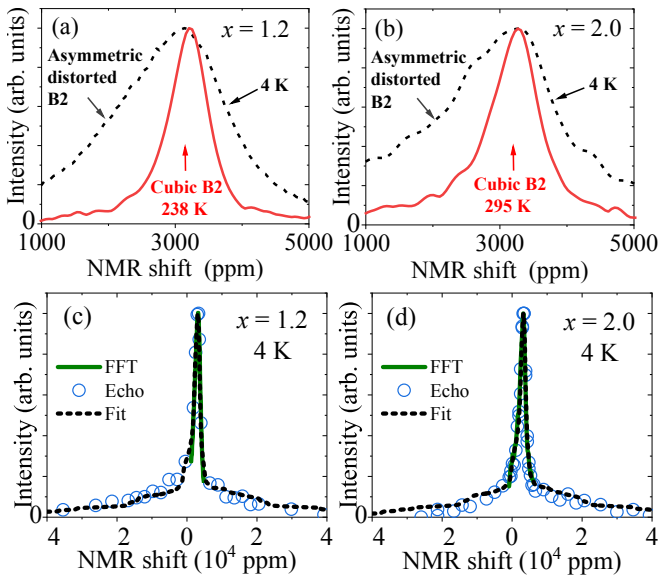

FIG. 6. (a,b) Ti NMR central-transition spectra for $x = 1.2$ and 2.0 strain-glass samples, relative to $^{47}$Ti standard. (c,d) Wide-line spectra, with fits described in the text.

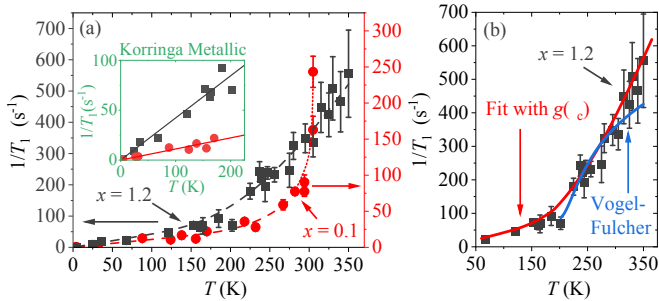

FIG. 7. (a) Fitted single-exponential $1/T_1$ for $x = 0.1$ and 1.2 samples versus temperature. Arrhenius plus Korringa fitted curves are shown by dashed curves. Korringa behaviors are shown by solid lines in the insert. Rapid increase in $T_1^{-1}$ above 300 K for $x = 0.1$ is shown by the dotted line, drawn as a guide to the eye. (b) Vogel-Fulcher and distributed-$\tau_c$ fits described in the text.

For a quantitative analysis of the enhanced $T_1^{-1}$ due to fluctuation, we consider Debye-type relaxation process with a single activated correlation time, equivalent to

Bloembergen, Purcell, and Pound theory [16, 47] denoted by $\left[\frac{1}{T_1}\right]_{\text{fluct}}$:

$$\left[\frac{1}{T_1}\right]_{\text{fluct}} = 4\pi^2 \nu_{\text{local}}^2 \frac{\tau_c}{1 + \omega_L^2 \tau_c^2}, \tag{1}$$

where $\tau_c^{-1} = f_0 \times \exp\left(-\frac{\Delta E}{k_B T}\right)$. Here, $\nu_{\text{local}}$ is the linewidth, $\Delta E$ is the activation energy, and $f_0$ is the attempt frequency, and $\omega_L = 2\pi\nu_L$. Under the assumption that $\nu_L^2 \tau_c^2 >> 1$, this simplifies to an Arrhenius formula:

$$\left[\frac{1}{T_1}\right]_{\text{fluct}} = f \exp\left(\frac{-\Delta E}{k_B T}\right), \tag{2}$$

where $f = f_0(\nu_{\text{local}}/\nu_L)^2$. As Fig. 7(a) shows, the experimental data is well fitted over the full temperature range for $x = 1.2$, and below the martensitic transition for $x = 0.1$, by the Arrhenius function plus the Korringa term $\left[\frac{1}{T_1}\right]_K = aT$:

$$\frac{1}{T_1} = \left[\frac{1}{T_1}\right]_{\text{fluct}} + \left[\frac{1}{T_1}\right]_K = f \exp\left(\frac{-\Delta E}{k_B T}\right) + aT. \tag{3}$$

The parameters were fitted to be $f = 2.2 \times 10^4$ $(5.9 \times 10^3)$ s$^{-1}$, $\Delta E = 0.155$ (0.087) eV, and $a = 0.11$ (0.42) (s K)$^{-1}$ for $x = 0.1$ (1.2). With $\nu_{\text{local}} = 300$ kHz ($\nu_Q$ for $^{49}$Ti which dominates the peak intensity region of the central transition), it is easy to show that $\nu_L \tau_c >> 1$ over the observed temperature range in both cases, validating the assumption for Arrhenius behavior. The observed attempt frequencies are much smaller than would be expected for single-atom type hopping, for which $f_0$ is typically on order of phonon frequencies, a few THz. Thus these results are consistent with macroscopic domain dynamics rather than that of localized defects or interstitials.

For $x = 0.1$, the results coincide with internal friction measurements [48], interpreted as due to the thermally activated motion of domain walls in the NiTi martensite. The internal friction results yielded $\Delta E = 0.29 \pm 0.02$ eV for this process, consistent with the result obtained here. There is a limited range for $x = 0.1$ Arrhenius fitting, nevertheless the results behave as expected. It is possible that the factor of $\lesssim 2$ difference is due to a non-Debye behavior for the domain wall relaxation function; a similar result is seen for NMR vs transport results in superionic conductors [49], however, it seems likely that domain wall motion of this type explains the NMR results.

For the $x = 1.2$ strain-glass, in the absence of well-defined martensite domains, the thermally activated behavior must instead be due to fluctuating disorder-induced strain. A Vogel-Fulcher-type thermally activated behavior has been reported in strain glass AC elastic measurements [4, 6, 8, 50], as well as other glass-forming systems [51–55], and is characterized by a modified thermally activated behavior, $\tau_c^{-1} = f_0 e^{-\Delta E/k(T-T_0)}$, where $T_0$ is typically close to the glass-freezing point. Vogel-Fulcher $T$-dependence has also been used to model the NMR $T_1$ for freezing processes in polymer glasses and other systems [56–58]. However, generalizing Eqn. [3] with the activation term replaced by $fe^{-\Delta E/k(T-T_0)}$, where $T_0$ was fixed to be 190 K, the fitted curve diverges from the data (Fig. 7(b)), indicating that Vogel-Fulcher behavior is not a good representation for the dynamics as a whole.

## V. DISCUSSION

### A. Local structure of strain glass

The large low-T linewidths in the strain glass samples can be understood as due to domains exhibiting strain which fluctuates at high temperatures thus causing motional narrowing of the NMR spectrum, but which becomes static in the low-temperature limit. Motional narrowing occurs when changes exceed the scale of the NMR linewidth, on the order of 100 kHz (equivalent to $\nu_{\text{local}}$ defined above). In this way, the apparent discrepancy with XRD results that are essentially T-independent can be understood, since the effective response time for XRD measurements is much shorter than that of NMR, so that the magnitudes of the strain within a given domain can appear unchanged in XRD at higher temperatures while in NMR their effect is diminished due to the motional-narrowing averaging process.

The large changes in NMR spectra (Fig. 3) coincide with the reported strain glass freezing temperatures of about 190 and 170 K for $x = 1.2$ and 2.0 based on previously reported phase diagram [5, 59, 60]. On the other hand, the magnetic shift exhibits very little change in the freezing process, e.g. for $x = 2.0$ for which $K$ extracted at 4 K is identical to the high-temperature shift. As $K$, the local susceptibility, is representative of the electronic behavior, this is an indication that the freezing process does not induce electronic changes, but only slows the dynamics of locally distorted regions.

For further information, we used the WIEN2K package to calculate the EFG's for stable and proposed phases in NiTi. For B19′ martensite, similar to previous report [40], we found that calculated EFGs are larger than what is measured in NMR, with $\nu_{\text{Q}}$ ranging from 980 to 1100 kHz for different reported B19′ structures [61–63], about 30% larger than the experimental 775 kHz. The strain-glass fitted results described above are comparable to these calculated B19′ values, while for the R phase (space group: $P3$) [63], we find somewhat larger calculated EFGs (see supplementary material [26]), with a weighted mean $\nu_{\text{Q}} = 1300$ kHz for the Ti sites. These results are both close to the experimental values for the strain glass materials, so that we can't obtain a precise indication of the local configuration, however taking the NMR and XRD results together, it is clear that the distortion of the local structure away from the cubic B2 configuration in the strain glass is significant, with a mean EFG on order of that of the B19′ and R structures as compared to B2 with zero EFG.

### B. Spin-lattice relaxation and metallic behavior

The low-temperature Korringa $T_1$ (Fig. 7(a)) indicates a large difference between the metallic behavior of the $x = 0.1$ sample and that of the strain glasses. The larger $1/(T_1T)$ for $x = 1.2$ corresponds to its significantly larger Fermi-level electronic density of states [$g(E_{\text{F}})$] than B19′ martensite. In addition, for $x = 2.0$ we measured $1/(T_1T)$ at the fixed temperature of 4 K. Since the fluctuation spectrum producing the enhanced $1/T_1$ at higher temperatures seems certain to have died out at 4 K as in the other two samples, we assume that the $x = 2.0$ result also represents its metallic behavior.

For a quantitative analysis, the approximate single-exponential recovery curve used above can be replaced by the exact multi-exponential central transition recovery curve for magnetic interactions, which for $I = 7/2$ is given by $M(t) = M(\infty)[1 - A(1.428e^{-\frac{28t}{T_1}} + 0.412e^{-\frac{15t}{T_1}} + 0.136e^{-\frac{6t}{T_1}} + 0.024e^{-\frac{t}{T_1}})]$. The $I = 7/2$ curve is appropriate since $^{49}$Ti dominates the NMR signal at the intensity peak, independent of EFG details. Refitted this way, we obtain $1/(T_1T) = 0.0052$, 0.017, and 0.014 (s K)$^{-1}$, for $x = 0.1$, 1.2, and 2.0, respectively, with $x = 1.2$ and 2.0 strain glasses similarly enhanced relative to the martensite.

A general expression for the Korringa $T_1$ [64, 65] is:

$$[\frac{1}{T_1T}]_{\text{K}} = 4\pi\hbar\gamma_{\text{n}}^2 k_{\text{B}} g_{\text{d}}^2(E_{\text{F}}) \sum_i \left(H_i^{\text{HF}}\right)^2 F_i, \quad (4)$$

where the sum extends over the hyperfine terms (orbital, core polarization, dipolar) for which $H_i^{\text{HF}}$ are the hyperfine fields, $\gamma_{\text{n}}$ is the nuclear gyromagnetic ratio, $g_{\text{d}}(E_{\text{F}})$ is the Ti-site local d-electron Fermi level density of states ($g_{\text{s}}(E_{\text{F}})$ and $g_{\text{p}}(E_{\text{F}})$ give negligible contributions relative to the d term), and $F_i$ is a numerical factor determined by the mixture of d orbitals [64, 65]. For core polarization we use the measured [66] $H_{\text{CP}}^{\text{HF}} = -12.6$ T, while $H_{\text{orb}}^{\text{HF}} = H_{\text{dip}}^{\text{HF}} = +\frac{\mu_0}{4\pi}2\mu_{\text{B}}\langle r^{-3}\rangle = 25.3$ T, where $\mu_0$ is the permittivity of free space and $\mu_B$ is the Bohr magneton, and the calculated $\langle r^{-3}\rangle = 1.36 \times 10^{31}$ m$^{-3}$ is used [67]. Note the hyperfine fields differ in sign, giving contributions to the Knight shift which tend to cancel, however for $T_1$, Eqn. (4) is a simple sum, with moreover an orbital term which is much larger than the other two, making the analysis unambiguous.

To analyze the results, we calculated $g_{\text{d}}(E_{\text{F}})$ resolved for individual orbitals, for the B2, B19′, and R phases, using the WIEN2K package. General relationships for $F_i$ are given in [65]; for the case that all 5 d orbitals are equally weighted at $E_{\text{F}}$, $F_{\text{orb}} + F_{\text{dip}} = 0.4 + 0.029 = 0.429$ and $F_{\text{CP}} = 0.2$, and we find for all structures the orbital mixtures give results close to these values; full details given in the supplementary information [26]. As a result, we obtained the calculated results $1/(T_1T) = 0.021$, 0.0037, and 0.015 (s K)$^{-1}$, for the B2, B19$'$ [62] and R [63] structures respectively. As the WIEN2k package computes local $g(E_F)$ inside the muffin-tin spheres only, and the additional interstitial contribution is about 20% of the total $g(E_F)$ in these phases, the calculated B19$'$ value is in good agreement with the measured result for $x = 0.1$, with the small observed value confirming the calculated pseudogap in $g(E_F)$ for B19$'$ as previously discussed [38, 68, 69]. On the other hand, the strain glass samples have considerably larger Korringa $1/(T_1 T)$, and hence larger $g_d(E_F)$ in the frozen state, with the observed $1/(T_1 T)$ in good agreement with that of the R phase.

## C. Nanodomain fluctuations

Returning to the enhanced $1/T_1$ observed at higher temperatures, the absence of a relaxation peak at the freezing temperature differs from the behavior typically found in relaxor ferroelectrics [46, 70, 71]. Similarly a peak in $1/T_1$ at the freezing temperature can be observed in some spin-glass systems [72]. In either case, the peak is indicative of a rapid decrease in the characteristic frequency of the fluctuation spectrum, crossing over to a quasistatic spectrum as the temperature crosses the freezing point. The contrasting behavior here indicates a much more gradual slowing down of the fluctuation spectrum over the entire temperature range.

The Arrhenius-type $1/T_1$ analysis described above gives us a probe of these dynamics. Based on the fitted $f = 5.9 \times 10^3$ s$^{-1}$ and $\Delta E = 0.087$ eV For $x = 1.2$ (Section IV C), at the freezing point of 190 K we obtain $f \exp\left(\frac{-\Delta E}{k_B T}\right) \cong 50$ s$^{-1}$ (Eqn. 3). Since $f = f_0(\nu_{local}/\nu_L)^2$ in Eqn. 3, and with $\nu_L = 21$ MHz and $\nu_{local} \cong {}^{49}\nu_Q \cong 300$ kHz (since $^{49}$Ti dominates the $T_1$ relaxation process), we obtain a correlation frequency $\tau_c^{-1} = f_0 \exp(-\Delta E/k_B T) = 200$ kHz at this temperature. As noted above the criterion for motional narrowing in NMR is that the correlation frequency crosses $\nu_{local}$ as it changes vs temperature. For Ti NMR, narrowing will commence when $1/\tau_c$ approached the central transition width at lower $T$, and proceed until $1/\tau_c$ exceeds $\mu_Q$. Since $\tau_c^{-1}$ obtained here from $1/T_1$ at the freezing point is indeed close to $\nu_{local}$, this shows the consistency between these two results and helps to further validate this model. Thus both the relaxation behavior and the observed NMR lineshape changes point to fluctuating local strain, with a fixed thermal activation barrier, as the process that dominates the local dynamics.

As an alternative model, we examine whether the NMR results can be fitted to a distribution of correlation times rather than a single thermally activated $\tau_c$. This has worked for other systems [73–76] and might account for the Vogel-Fulcher behavior observed in the mechanical relaxation of spin glass systems [6, 77]. Defining $g(\tau_c)$, as the distribution function representing the spatial variation of the fluctuation spectrum, we replace Eqn. 1 by the corresponding integration:

$$\left[\frac{1}{T_1}\right]_{fluct} = \int g(\tau_c) d\tau_c \frac{4\pi^2 \nu_{local}^2 \tau_c}{1 + \omega_L^2 \tau_c^2}. \tag{5}$$

Since the NMR results point to an Arrhenius spectrum, we assume a distribution weighted toward small $\tau_c$ for which the temperature dependence is activated, with a large $\tau_c$ tail which has a Vogel-Fulcher divergence above the freezing temperature. This is equivalent to a model previously developed for the dielectric behavior of relaxor ferroelectrics [75]. Accordingly, we assigned $\tau_1$ and $\tau_2$, respectively, as small and large $\tau_c$ cutoffs, with the fitted [75] distribution $g(\ln \tau_c) \propto \ln(\tau_2/\tau_c)$, (or equivalently $g(\tau_c) \propto \left[\ln\left(\frac{\tau_2}{\tau_c}\right)\right]/\tau_c$ as plotted in Fig. 8), appropriately normalized.

We set $\tau_1 = \tau_{1,0} e^{1000\ \text{K}/T}$, with activation energy $k_B \times 1000$ K $= 0.087$ eV to match the NMR relaxation results, with the $\tau_{1,0}$ parameter adjusted to fit $1/T_1$, while for $\tau_2$ we use parameters based on internal friction in a NiTi strain glass [6], $\tau_2 = \tau_{2,0} e^{[150\ \text{K}/(T - T_0)]}$, with $T_0 = 160$ K and $\tau_{2,0} = 6 \times 10^{-5}$ s. Below the glass freezing temperature (190 K for $x = 1.2$), however, the apparent critical divergence of the largest $\tau_c$ clusters is interrupted, presumably as they become constrained by neighboring highly-pinned regions. To make $\tau_c$ continuous versus $T$ we set $\tau_2 = \tau_{2,0} e^{950\ \text{K}/T}$ below 190 K, with the same attempt frequency but modified activation energy due to the arrested divergence.

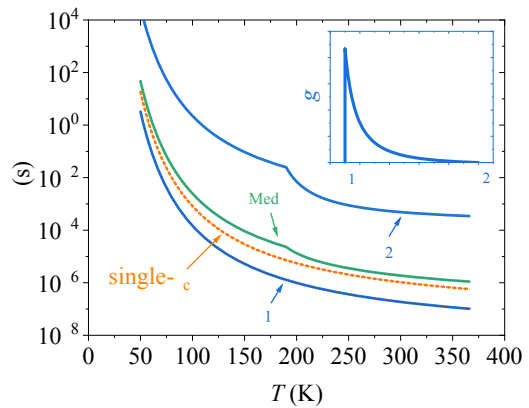

FIG. 8. Limiting parameters for fitted $\tau_c$ distribution: $\tau_1, \tau_2$ end points and median value ($\tau_{Med}$). Also single-$\tau_c$ result equivalent to narrow distribution. Inset: distribution function $g(\tau)$ for case that $\tau_2/\tau_1 = 10$.

Fitting the above described model to the $1/T_1$ results (see Fig. 7(b)), gives $\tau_{1,0} = 7 \times 10^{-9}$ s. This yields good agreement, similar to the single-$\tau_c$ result. The solid curves in Fig. 8 show the corresponding temperature dependence for $\tau_1, \tau_2$, with the middle solid curve also showing the median value, $\tau_c \equiv \tau_{Med}$ for this distribution, and the dashed curve showing the fitted dynamics based on the single-$\tau_c$ fitting. At 130 K for example, the latter gives $1/\tau_c = 12$kHz, equivalent to 600 ppm, roughly equal to the central-transition width. At this temper-ature the linewidth is about half-way between the low-temperature width and the high-temperature narrowed value (Fig. 3(d)) so that as described above these are reasonable values to explain the gradual narrowing process in this temperature regime. At this temperature the distribution has $1/\tau_{Med} = 4$ kHz (200 ppm), somewhat smaller than the central-transition width. This suggests that a more complicated distribution function in which the median cluster is unaffected by the divergence would better model this system, however this distribution captures the general features, considering that the median represents the typical cluster so that the progressive decrease of $\tau_{Med}$ versus $T$ implies that a progressively larger fraction of the line becomes narrowed, as observed.

Our results indicate that the Vogel-Fulcher divergence represents a very small fraction of the strain glass in this system, with the majority of the material retaining the relatively fast dynamics required to explain the NMR spectra, extending well below the freezing temperature. This is therefore distinct from the behavior of spin glasses which appear to freeze in a more uniform way [73]. Also the distribution of dynamical behavior obtained here is more skewed that the distribution which was fitted to the TiPdCr system [76]. We also point out that high frequency ultrasonic measurements in the TiPdCr strain glass appear to deviate from a Vogel-Fulcher relation [78], as do extended-frequency anelastic studies of $Ni_{50+x}Ti_{50-x}$ alloys [79], results which are consistent with non-divergent behavior of the fluctuation spectrum for small correlation times (e.g. large frequencies) as indicated here. However our model shows that a low-frequency Vogel-Fulcher type tail can at the same time be consistent with the storage-modulus results from AC elastic measurements.

## VI. CONCLUSIONS

We show that NMR can clearly indicate the freezing process in $Ni_{50+x}Ti_{50-x}$, applying this probe for the first time to a strain glass system. The results indicate the development of static local distortions at low temperatures as a slowing-down process, revealing significant local strains in the frozen configuration, with large nuclear electric field gradients comparable to that of the martensite phase despite the absence of a displacive phase transformation. The dynamical unfrozen strain glass causes a large enhancement of the NMR spin-lattice relaxation rate, from which we conclude that the freezing process involves predominantly an Arrhenius process above and below the glass-freezing temperature. From the further analysis we present a model for the locally inhomogeneous dynamical strain, which includes a very small density of critically divergent domains involved in the Vogel-Fulcher process, but strongly skewed toward faster dynamics involving more strongly pinned domains.

## VII. ACKNOWLEDGMENTS

Support for this project was provided by the Texas A&M University T3 program. Use of the Advanced Photon Source at Argonne National Laboratory was supported by the U. S. Department of Energy, Office of Science, Office of Basic Energy Sciences, under Contract No. DE-AC02-06CH11357. IK and DS acknowledge the support from The U.S. National Science Foundation, Division of Materials Research, Grant No. 1905325.

*Supplemental material for:*

# NMR study of Ni$_{50+x}$Ti$_{50-x}$ Strain-Glass

*Rui Li, Jacob Santiago, Daniel Salas, Ibrahim Karaman, and Joseph H. Ross Jr.*

## 1) X-ray measurements

Further x-ray diffraction (XRD) details are as follows:

**a) Results for $x = 0.1$ sample:** For the $x = 0.1$ sample with the conventional martensite phase transition, the cubic B2 lattice constant refined to 3.0205 Å at 400 K. Aside from 0.8% per mole formula-unit Ti$_2$Ni, no TiC or other phases were detected [1]. For 100 K, fig S1(a) shows the refinement described in the text fitted to the B19′ structure [2] combined with the small Ti$_2$Ni reflections. The structure given in reference [3] gave nearly identical results once the atomic parameters were relaxed. The small fitted lattice constant changes for 100 K relative to the reported low temperature values [2] are consistent with the reported anisotropic thermal expansion of the B19′ phase [4]. The broad intensity peak near 12° in the difference curve has no features that could be matched to a known phase, although the position corresponds to the region of largest intensity for the R phase. Assuming this possibility, an added R phase component [5] led to a refined value of 8% phase fraction, but only with the addition of very large microstrain broadening corresponding to 4.5% RMS strain for the R phase. This improved the $R_w$ from 15.9% to 12.5% [Fig. S1(b)]. A similar fit was obtained assuming heavily size broadened R phase (refined to 7 nm domain size), and a nearly equivalent fit could be obtained with a heavily broadened orthorhombic B19 phase [6], making it unclear what might be the significance of this feature.

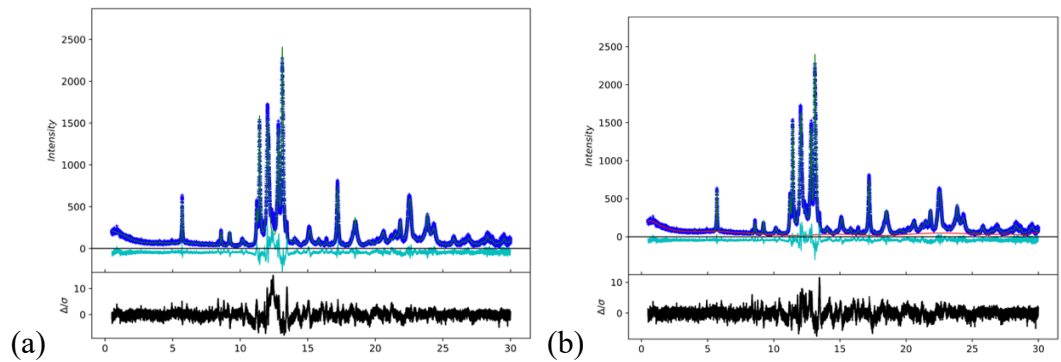

Figure S1: XRD refinement results for $x = 0.1$ (Ni$_{50.1}$Ti$_{49.9}$) sample at 100 K. (a) B19′ phase [2] plus 0.8 % Ti$_2$Ni. Refined lattice constants $a = 2.8804$ Å, $b = 4.1152$ Å, $c = 4.6543$ Å, $\beta = 97.58°$. (b) with heavily strain-broadened R phase as described in text.

**b) Results for $x = 1.2$ and 2.0 strain glass samples:** For the strain glass samples, as described in the main text, all spectra were fitted to a model having reflections associated with a microstrain-broadened B2 phase overlaying an identical B2 phase with narrow reflection profiles. Results for 200 K and 295 K are shown in Fig. S2, while the 100 K results are shown in the main text. For $x = 1.2$, refinement yielded 26% phase fraction for the broadened B2 phase, with 1.6% microstrain at 295 K increasing to 2.2% at 100 K. For *x* = 2.0 the results yielded 1.9% microstrain/48% phase fraction changing to 1.9%/41% at 100 K.

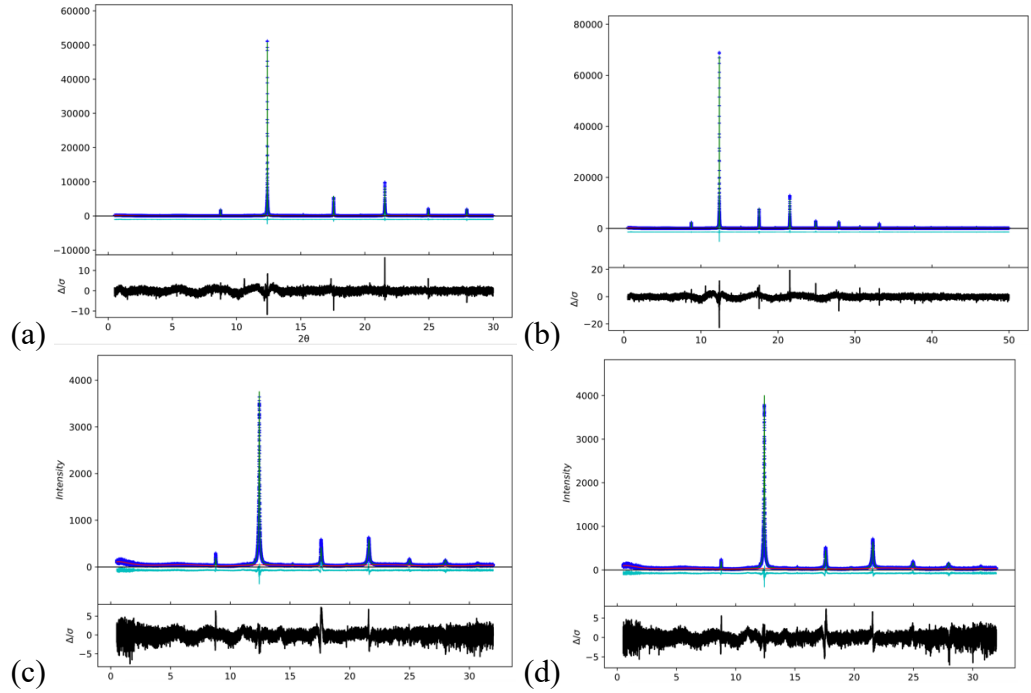

<u>Figure S2</u>: Refined Ni$_{50.1}$Ti$_{50-x}$ XRD results for (a) x = 1.2 at 200 K ($R_w$ = 8.6%); (b) x = 1.2 at 295 K ($R_w$ = 9.7%); (c) x = 2.0 at 200 K ($R_w$ = 14.3%); (d) x = 2.0 at 295 K ($R_w$ = 14.1%).

A further comparison is shown in Fig. S3, for which R phase [5], with 10% phase fraction per mole formula unit and the same peak profile as the B2 phase, has been added to the refined curves for *T* = 100 K. It is clear that the R phase is not represented in the data, as compared to the reported results [7] after a prolonged soaking time.

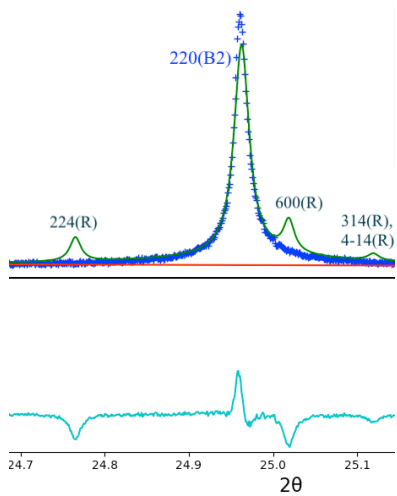

<u>Figure S3</u>: XRD refinement for *x* = 1.2 sample at 100 K, fitted to B2 phase with added R phase (10% R-phase fraction) added to the calculated curves. Results shown in the vicinity of the (220) B2 reflection, with individual calculated reflections labeled.

The fitted B2-phase lattice constants for the $Ni_{50+x}Ti_{50-x}$ strain glass materials are shown in Fig. S4. The results obtained at 200 and 295 K correspond to a linear coefficient of thermal expansion $\beta = 11$ and $9.1 \times 10^{-6}$/K for $x = 1.2$ and 2.0 respectively. These can be compared to values in the range $12 - 13 \times 10^{-6}$/K measured for the B2 phase in equiatomic NiTi at higher temperatures [8–10]. Extending to lower temperatures, one can expect a somewhat nonlinear behavior according to the thermodynamic equality $\beta = \frac{\gamma C_V}{9B}$ [11,12], where $C_V$ is the specific heat and $B$ is the bulk modulus, with the Grüneisen constant $\gamma$ taking into account the lattice anharmonicity. Taking the calculated [13] $B = 150$ GPa for the B2 phase, and with the Debye temperature taken to be $\Theta_D = 290$ K [14], we obtained the curves $a = a(0)[1 + \int \beta \, dT]$ shown in the figure. These are based on a Debye function for $C_V$, assuming $B$ and $\gamma$ to be temperature independent. The least-squares fit gives $a(0) = 3.0025$ (2.9967) Å for $x = 1.2$ (2.0), and corresponding Grüneisen constants $\gamma = 1.6$ (1.35). These estimated values for $\gamma$ are in line with typical values for metals, and though there may be some composition dependence for $B$ and $\Theta_D$, there is no indication that the results depart from typical thermal-expansion behavior, even though the strain-glass freezing temperatures fall in the middle of the range. Thus similar to the xrd profiles described above, which exhibit very little change going into the strain glass phase, the lattice constants also display no obvious signs of anomalous behavior when sampled across the spin glass freezing temperature.

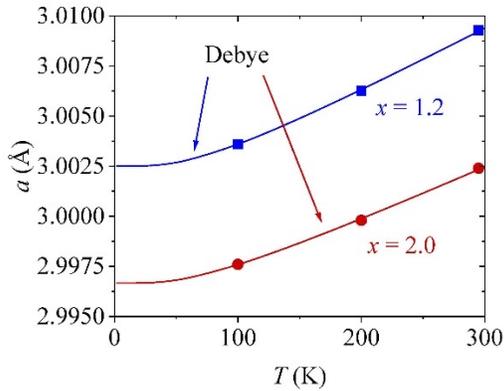

Figure S4: Lattice constants for B2 phase obtained from XRD refinements for $Ni_{50.1}Ti_{50-x}$ strain glass samples x = 1.2 and x = 2.0, along with Debye-model thermal expansion curves described in the text.

## 2) DFT calculations: Electric Field Gradients and Korringa $T_1$

DFT calculations were performed using the Wien2k package, with more details given in the main text. Partial densities of states are shown in Figures S5 and S6 for the NiTi B19′, B2, and R phases, and Fermi-level parameters are given in Table S1.

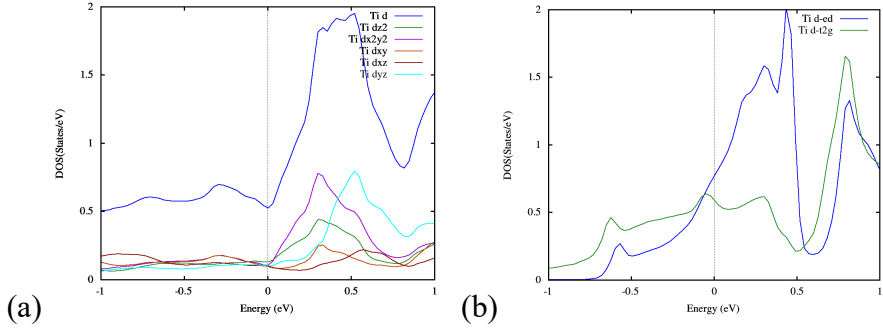

Figure S5: Ti partial densities of states near the Fermi energy for NiTi in (a) B19' structure with structural parameters from [3]. Blue curve: total Ti-$g_d(E)$. Other curves: $g_d(E)$ for individual d orbitals as shown. (b) B2 structure with a = 3.015 Å. Blue: $e_g$ symmetry; Green: $t_{2g}$ symmetry.

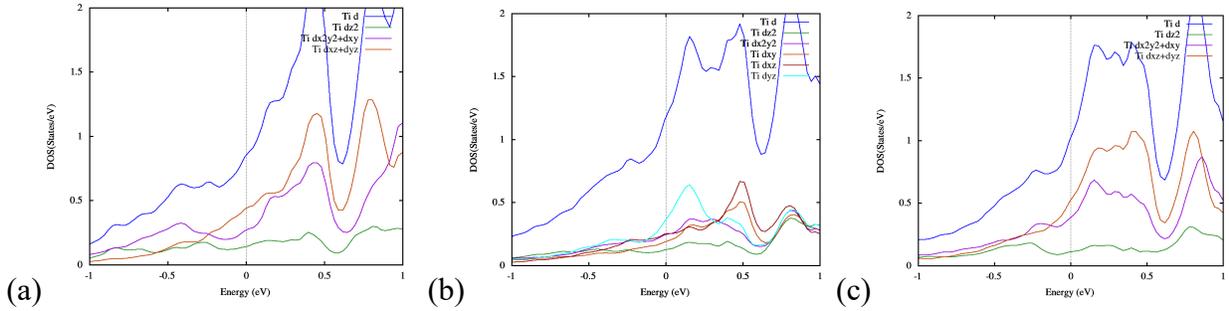

Figure S6: Ti partial densities of states near the Fermi energy for NiTi R-phase, parameters from [5]: (a) site 1; (b) site 2; (c) site 3. Blue curves: total Ti-$g_d(E)$. Other curves for orbitals indicated.

For Korringa-type spin-lattice relaxation in NMR, as given in the main text, a general expression is,

$$\frac{1}{T_1 T} = 4\pi\hbar\gamma_n^2 k_B g_d^2(E_F) \sum_i (H_i^{HF})^2 F_i, \qquad [S1]$$

where $i$ refers to individual hyperfine terms, with orbital (*orb*), core polarization (*CP*), and dipolar (*dip*) relevant here and $H_i^{HF}$ is the hyperfine field. Also $\gamma_n$ is the nuclear gyromagnetic ratio, $g_d(E_F)$ is the Ti-site local $d$-electron Fermi level density of states. Expressions for the numerical factors $F_i$ are given for general symmetry in [15], depending on the mixture of $d$ orbitals at $E_F$. Since both $H_{orb}^{HF}$ and $H_{dip}^{HF}$ are equal to $\frac{\mu_0}{4\pi} 2\mu_B \langle r^{-3} \rangle$, we report the orbital and dipolar $F_i$ values together in the table. For core polarization, $F_{CP} = \lambda_{z^2}^2 + \lambda_{x^2-y^2}^2 + \lambda_{xy}^2 + \lambda_{yz}^2 + \lambda_{xz}^2$, with $\lambda_j$ representing the fraction of $g_d(E_F)$ due to the $j^{th}$ orbital, so that $\sum \lambda_j = 1$. For all of the phases considered here, the Ti $d$-orbitals contribute in a relatively uniform way, so that $F_{CP}$ is close to the minimum value of 0.2 in all cases (Table S1). Along with the larger $H_{orb}^{HF}$ compared to $H_{CP}^{HF}$, the orbital contribution to $1/T_1T$ thus turns out to be significantly larger than the others. Taken together, these factors along with Equation [S1] yield the calculated Korringa $1/T_1T$ values shown in the table.

Table S1: Calculated local densities of states, Ti-site hyperfine parameters, and Ti-NMR Korringa-$1/(T_1T)$, and EFG parameters for NiTi structures as shown. Factors $F_{orb}$, $F_{dip}$, $F_{CP}$ for orbital, dipolar, and core polarization hyperfine interactions are calculated using general relationships given in [15].

| structure | multiplicity | $g_{total}(E_F)$ states/NiTi unit/eV | $g_d(E_F)$ $d$-states/Ti atom/eV | $F_{orb} + F_{dip}$ | $F_{CP}$ | calculated $1/(T_1T)$ [1/(sK)] | $^{47}\nu_Q$ [kHz] | $\eta$ |
|---|---|---|---|---|---|---|---|---|
| B2  (a) | 1 | 3.182 | 1.350 | 0.371 | 0.222 | 0.0215 | 0 | – |
| B19' (b) | 2 | 1.478 | 0.547 | 0.426 | 0.203 | 0.0037 | 1080 | 0.99 |
| R (c): | | | | | | | | |
|   site 1 | 1 | | 0.856 | 0.422 | 0.210 | 0.0090 | 1700 | 0 |
|   site 2 | 6 | | 1.180 | 0.405 | 0.214 | 0.0165 | –1100 | 0.96 |
|   site 3 | 2 | | 1.024 | 0.409 | 0.215 | 0.0126 | 1620 | 0 |
|   site-weighted average | | 2.447 | 1.109 | | | 0.0148 | 1280 | – |

(a) Cubic B2 phase with $a$ = 3.015 Å.
(b) B19' parameters from reference [3].
(c) R phase parameters from reference [5].